\documentclass[10pt,conference]{IEEEtran}
\IEEEoverridecommandlockouts

\usepackage[T1]{fontenc}
\usepackage[utf8]{inputenc}
\usepackage{microtype}

\usepackage{amsmath,amssymb}
\usepackage{graphicx}
\usepackage{booktabs}
\usepackage[ruled,vlined,linesnumbered]{algorithm2e}
\usepackage{tikz}
\usetikzlibrary{positioning,arrows.meta,fit,calc,backgrounds}

\usepackage{listings}
\usepackage{xcolor}
\newif\ifshowchanges\showchangesfalse
\newenvironment{chg}{\ifshowchanges\color{blue}\fi}{}
\newcommand{\chgi}[1]{{\ifshowchanges\color{blue}\fi#1}}

\definecolor{kw}{RGB}{0,0,180}
\definecolor{cmt}{RGB}{106,115,125}
\definecolor{str}{RGB}{170,55,60}
\definecolor{num}{RGB}{0,128,128}
\lstdefinestyle{py}{
  language=Python,
  basicstyle=\ttfamily\small,
  keywordstyle=\color{kw}\bfseries,
  commentstyle=\color{cmt}\itshape,
  stringstyle=\color{str},
  numberstyle=\color{num},
  showstringspaces=false,
  extendedchars=true,
  frame=single, framesep=4pt, xleftmargin=4pt,
  aboveskip=3pt, belowskip=3pt,
  literate={²}{{\textsuperscript{2}}}1   %
}

\usepackage{cite}
\usepackage{url}
\usepackage[hidelinks]{hyperref}  %
\usepackage[capitalize]{cleveref}  %

\newcommand{\anon}[1]{\textcolor{red}{[anonymized]}}

\newif\ifcomments\commentsfalse  %

\newcommand{\numRepos}{11}            %
\newcommand{\totalLOC}{1,156,044}
\newcommand{\numContracted}{5,406}    %
\newcommand{\pctExecuted}{81}         %
\newcommand{\pctExecutedSetup}{29.5}  %
\newcommand{\pctExecutedLite}{4.5}    %
\newcommand{\numReproducedBugs}{1,050}%
\newcommand{\numSpecBugs}{829}        %
\newcommand{\numCHBugs}{221}          %
\newcommand{\numEndToEndBugs}{920}    %
\newcommand{\pctEndToEnd}{88}         %
\newcommand{\numBaselineBugs}{1,824}  %
\newcommand{\numAdditionalBugs}{312}  %
\newcommand{\pctImprovement}{17}      %

\begin{document}

\title{Bug Finding with Agentic Symbolic Execution\thanks{Preprint under review.}}

\author{\IEEEauthorblockN{George Granberry, Wolfgang Ahrendt, Moa Johansson}
\IEEEauthorblockA{\textit{Chalmers University of Technology}\\
Gothenburg, Sweden\\
\{georgegr, ahrendt, moa.johansson\}@chalmers.se}
\and
\IEEEauthorblockN{Andrew D. Gordon}
\IEEEauthorblockA{\textit{University of Edinburgh}\\
Edinburgh, United Kingdom\\
andy.gordon@ed.ac.uk}}

\maketitle

\begin{abstract}

Symbolic execution, in tools such as CrossHair for Python, is a powerful technique for finding bugs, but two barriers hinder its adoption: the human effort of annotating code with assertions, and the need to model the execution environment outside Python, such as an external database.
To overcome both, we design \chgi{an agentic harness that orchestrates the symbolic execution tool CrossHair together with an AI coding agent}.
Without human intervention, it instruments a repository with assertions, synthesizes a constraint-aware symbolic environment for the database, runs symbolic execution, and constructs a concrete test as evidence for each bug it reports.
We evaluate our harness on \totalLOC\ lines of Python code over \numRepos\ popular open-source Python web applications (Zulip, NetBox, Apache \chgi{Superset}, and others).
\begin{chg}
With the synthesized environment, CrossHair executes \pctExecuted\% of the \numContracted\ contracted functions, against \pctExecutedLite\% without it.
Our pipeline finds \numReproducedBugs\ bugs, each reproduced by a test against the unmodified application, \numCHBugs\ of them by CrossHair, and \numAdditionalBugs\ of them are missed by a baseline in which the same LLM inspects the code.
The two find different kinds of bugs: CrossHair's are mostly crashes on input-domain corner cases that validation admits, such as empty strings, missing fields and unusual characters, while inspection's are mostly bugs that produce wrong output. Our framework thus lets symbolic execution supplement inspection-based bug finding.
\end{chg}
\end{abstract}


\section{Introduction}
\label{sec:introduction}

\begin{figure*}[t]
  \centering
  \resizebox{\textwidth}{!}{%
  \begin{tikzpicture}[
    font=\small,
    >={Stealth[round]},
    stage/.style={rectangle, rounded corners=2pt, draw=blue!55!black, fill=blue!10,
                  minimum height=9mm, minimum width=16mm, align=center, inner sep=2pt},
    gate/.style={rectangle, rounded corners=2pt, draw=orange!75!black, fill=orange!18,
                 minimum height=9mm, minimum width=14mm, align=center, inner sep=2pt},
    io/.style={rectangle, rounded corners=7pt, draw=black!70, fill=black!3, align=center,
               minimum height=8mm, minimum width=14mm, inner sep=2pt},
    band/.style={rectangle, rounded corners=3pt, draw=olive!70!black, fill=yellow!12,
                 minimum height=8mm, align=center, inner sep=3pt},
    codebox/.style={rectangle, rounded corners=1.5pt, draw=gray!50, fill=gray!5,
                    font=\ttfamily\scriptsize, align=left, inner sep=3pt, line width=0.4pt},
    flow/.style={->, thick},
    feed/.style={->, semithick, gray!30!black},
    drive/.style={->, dashed, gray!55!black},
    repair/.style={->, dashed, orange!75!black, thick},
    elab/.style={font=\scriptsize, inner sep=1pt},
    lbl/.style={font=\scriptsize, fill=white, inner sep=1.5pt},
    num/.style={circle, draw=black!75, fill=white, font=\scriptsize\bfseries,
                inner sep=0.6pt, minimum size=3.6mm},
  ]
    \node[codebox] (src) at (-0.6,0) {%
      \textcolor{gray!55!black}{\# models.py}\\
      \textcolor{blue!60!black}{class} Currency(Model):\\
      \ \ rate = DecimalField()\\
      \ \ \textcolor{blue!60!black}{def} save(self, *a):\\
      \ \ \ \ super().save(*a)};
    \coordinate (fork) at ($(src.east)+(0.25,0)$);
    \node[stage]   (s1)   at (2.4,1.5)  {Environment\\Synthesis\\\chgi{\scriptsize(stubs + lazy init)}};
    \node[codebox] (ir)   at (4.85,1.5)  {%
      \textcolor{gray!55!black}{\# schema IR}\\
      Currency.rate:\\
      \ max\_digits=10\\
      \ null=False};
    \node[codebox] (stub) at (7.3,1.5)  {%
      \textcolor{gray!55!black}{\# stubs.py}\\
      Currency.save =\\
      \ \ \ \textcolor{blue!60!black}{mock\_save}\\
      add(rate >= 0)};
    \node[stage]   (s2)   at (2.4,-1.5) {Contract\\Generation};
    \node[codebox] (con)  at (4.9,-1.5) {%
      \textcolor{gray!55!black}{\# models.py}\\
      \textcolor{blue!60!black}{def} save(self):\\
      \ \textcolor{orange!80!black}{"""post:}\\
      \ \ \textcolor{orange!80!black}{rate >= 0"""}};
    \node[stage] (s3)     at (9.5,0)  {Symbolic\\Execution};
    \node[gate]  (s4)     at (11.5,0) {Audit};
    \node[stage] (s5)     at (13.9,0) {Witness\\Validation};
    \node[io]    (report) at (15.8,0) {Report};
    \draw[flow] (src.east) -- (fork);
    \draw[flow] (fork) to[out=70,in=180]  (s1.west);
    \draw[flow] (fork) to[out=-70,in=180] (s2.west);
    \draw[feed] (s1) -- (ir);
    \draw[feed] (ir) -- (stub);
    \draw[feed] (s2) -- (con);
    \draw[flow] (stub.east) to[out=0,in=180] ([yshift=2mm]s3.west);
    \draw[flow] (con.east) to[out=0,in=180]
       node[lbl,pos=0.5] {contracted function} ([yshift=-2mm]s3.west);
    \draw[flow] (s3) -- (s4);
    \draw[flow] (s4) -- node[elab,above] {accepted} (s5);
    \draw[flow] (s5) -- (report);
    \draw[feed] (ir.south) to[out=-100,in=30]
       node[lbl,pos=0.5] {schema} (s2.north east);
    \draw[flow] (s2.south) -- (2.4,-2.75) -- node[lbl,pos=0.5] {flagged defects}
       (11.5,-2.75) -- (s4.south);
    \foreach \st/\n in {s1/1,s2/2,s3/3,s4/4,s5/5}{
      \node[num] at (\st.north west) {\n};
    }
    \node[band, minimum width=13cm] (agent) at (7.6,2.9) {agentic harness + skill library};
    \foreach \st/\verb in {s1/generate,s3/run,s4/judge,s5/validate}{
      \draw[drive] (agent.south -| \st.north) -- (\st.north)
        node[elab, midway, right=1pt] {\verb};
    }
    \draw[drive] (1.35,2.45) -- (1.35,-1.5) -- (s2.west);
    
  \end{tikzpicture}%
  }
  \caption{System overview.}
  \label{fig:system-overview}
\end{figure*}
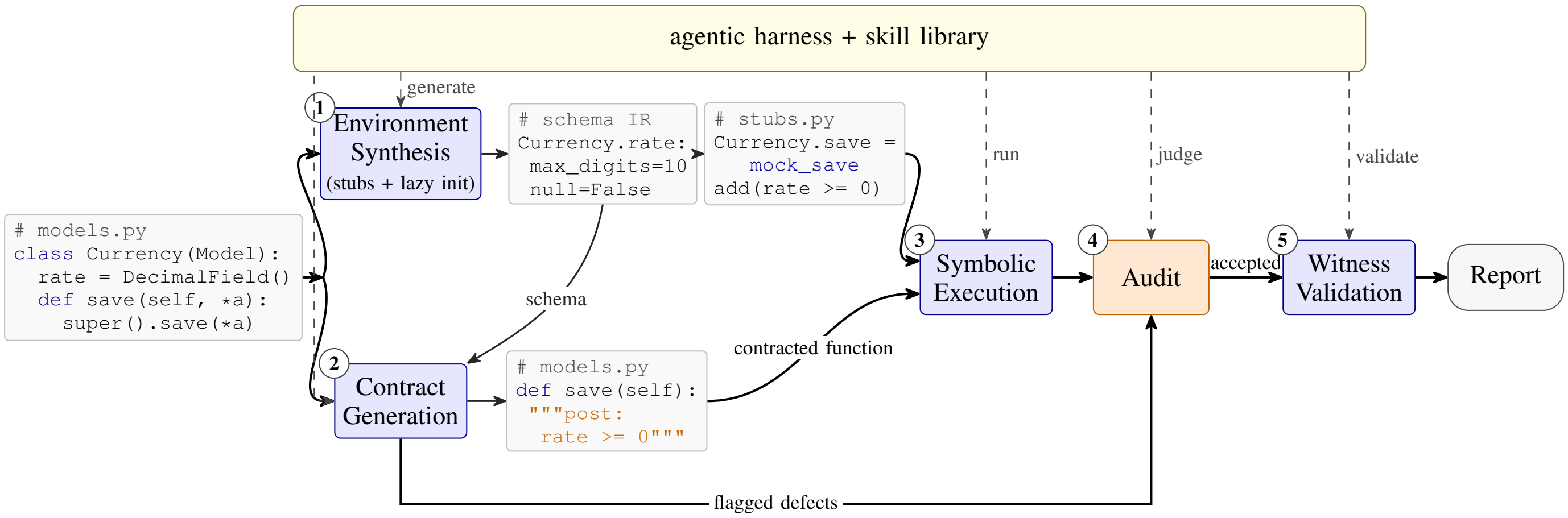

\begin{chg}
Contracts (pre- and postconditions) let a symbolic execution engine check code
automatically: the engine propagates constraints along each path and asks an SMT
solver for an input that violates a contract, which it returns as a
counterexample. Two things stand in the way on real applications. Their code
ships without contracts, and code that calls into an external system such as a
database cannot run symbolically without a model of that system.
CrossHair~\cite{schanely2017crosshair}, a symbolic execution engine for Python,
needs both. \emph{Our work makes symbolic execution available, with full
automation, to code that has neither contracts nor models of its external
calls.}

Our neuro-symbolic agent \emph{crosshair-bugs} combines LLM-based specification
generation with symbolic execution. The agent writes a contract for each
function and a symbolic model of the database the function touches, and
CrossHair then searches for inputs that violate the contracts. Every resulting
bug candidate is judged by a separate audit agent, which attributes it to the
application, to the generated contract, or to the symbolic environment.
Candidates judged to be application bugs are turned into test cases, validated
as witnesses by \emph{concrete} execution, and reported to the user.

We target database-backed web applications written in Python, a popular class
of software whose behavior depends on external calls and which symbolic
analysis has therefore rarely reached. We evaluate on \totalLOC\ lines of
Python in \numRepos\ popular open-source applications (Zulip, NetBox,
Apache Superset, and others), against a purely LLM-based baseline, and ask:
\begin{enumerate}
  \item \textbf{RQ1 (Reach):} What fraction of the contracted functions can
    symbolic execution execute with agentic environment synthesis, compared to
    without it?
  \item \textbf{RQ2 (Exploration depth):} How deeply does symbolic execution
    explore the functions it executes, and what does it report?
  \item \textbf{RQ3 (Bugs):} What bugs does symbolic execution find in this
    code, and which of them does direct LLM inspection miss?
\end{enumerate}

\paragraph{The need for hand-written models}
Symbolic-execution research has long had answers to the problems that stop an
engine on real code: environment models and mocks for calls the engine cannot
follow~\cite{dehalleux2010moles,priya2024vmocks}, symbolic treatment of
database state~\cite{emmi2007dynamic,pan2013guided,marcozzi2015relational},
and lazy initialization for inputs that are complex
objects~\cite{khurshid2003generalized,rosner2015bliss,ramos2015under}. Each
must be instantiated for every project: someone has to write the stubs, the
invariants that bound a lazily built object, and the model of the external
system. This per-project manual work is a large part of why these techniques
have rarely reached large database-backed applications.

\paragraph{Our idea}
We use an agent as the missing instantiation layer. Rather than taking an
environment model as input, our pipeline produces it, through
\emph{knowledge-seeded environment synthesis}: the model is seeded by framework
knowledge packaged in the agent's skills and by the agent's own inspection of
the repository, and checked by loading it into CrossHair. The same agent writes
the contracts that serve as the oracle, and executable witnesses against the
real application discharge whatever the over-approximate model admits.

\paragraph{Findings}
\emph{RQ1} (\cref{sec:results-rq1,fig:rq1-reach}): bare CrossHair executes
\pctExecutedLite\% of the \numContracted\ contracted functions, and the
framework initialization a test suite performs \pctExecutedSetup\%. With the
synthesized environment it executes \pctExecuted\%.
\emph{RQ2} (\cref{sec:results-rq2,fig:rq2}): 28.6\% of the functions are
explored to at least 25 paths without a violation, and 3.6\% exhaustively.
39.1\% are refuted, but most refutations are false positives of the
over-approximate environment, which the audit removes.
\emph{RQ3} (\cref{sec:results-rq3,fig:rq3-discovery,fig:rq3-overlap}): the
pipeline finds \numReproducedBugs\ bugs, each reproduced against the unmodified
application and \numEndToEndBugs\ of them end-to-end: \numCHBugs\ found by
CrossHair and \numSpecBugs\ flagged while writing contracts. A baseline that
asks the same model to inspect the code reproduces more bugs
(\numBaselineBugs) but misses \numAdditionalBugs\ of ours, including a third
of CrossHair's. The two effects meet in 110 of CrossHair's bugs, which lie in
functions that execute only under the synthesized environment. The baseline
misses 40 of these, so neither tool alone would have found them.

\noindent Our contributions are:
\begin{itemize}
\item \emph{Knowledge-seeded environment synthesis} for ORM
  call sites: an agent builds a symbolic model of the database from framework
  knowledge and repository inspection, and repairs it until CrossHair can
  load it (\cref{subsec:env-modeling}).
\item \emph{Agent-instantiated symbolic-execution techniques}: schema-constrained
  ORM stubs, and bounded lazy initialization whose bounds the agent infers from
  the code, installed through CrossHair's plugin interface without modifying
  the engine (\cref{subsec:env-modeling}).
\item A design principle, \emph{the LLM proposes, the solver checks, concrete
  execution confirms}, realized as a reproducible agent harness
  (\cref{subsec:sys,subsec:harness}).
\item An evaluation on \totalLOC\ lines of Python in \numRepos\ web
  applications, reporting reach, exploration depth, and the
  \numCHBugs\ witness-reproduced bugs symbolic execution finds, compared
  against a same-model, witness-validated LLM baseline.
\end{itemize}

\paragraph{Scope of the claim}
We do not argue that symbolic execution finds more or more important bugs than
an LLM reading the code: on our corpus it does not. We show that, once an agent
supplies the environment, symbolic execution runs on large, complex
database-backed applications at scale and finds the bugs it is known for:
input-domain corner cases that validation admits and that LLM inspection with
the same model often misses.
\end{chg}

\section{Background}
\label{sec:background}

\subsection{CrossHair}
\label{subsec:bg-crosshair}
CrossHair is a concolic~\cite{sen2005cute,godefroid2005dart} analysis tool for Python that searches for inputs
that violate a function's specification~\cite{schanely2017crosshair}. Specifications are supplied as contracts seen in Fig.~\ref{fig:pure-vs-orm} and CrossHair attempts to find a concrete input, admitted by the pre-conditions, for which an exception is
raised or a post-condition evaluates to false. Such an input is reported as a
counterexample.

CrossHair calls
the function under test, substituting \emph{symbolic} proxy objects for its
arguments. A parameter annotated \texttt{int}, for instance, is supplied as a
\texttt{SymbolicInt} that carries, instead of a concrete value, an SMT-LIB~\cite{smtlib} term of sort \texttt{Int} over symbolic variables, which CrossHair builds through the API of the SMT solver Z3~\cite{z3}.

As the function executes, operations on these symbolic types build up
Z3 constraints, and the branch conditions encountered along a path are
accumulated into a path condition. CrossHair then uses Z3 to identify feasible paths, and records its decisions so that subsequent executions explore alternative branches. When a target
condition is reached, it queries the solver for a variable assignment satisfying the precondition and path conditions, but violating the target condition, and surfaces it to the user. CrossHair is consequently effective at discovering
counterexamples and bringing higher assurance than the typical test. It does not, however, offer a guarantee of the absence of a counterexample.

A contract constrains a single function, but CrossHair's behavior at function
boundaries follows directly from this execution model and governs how a
misstated contract behaves. A callee's \emph{precondition} is enforced at the
call site: invoking a contracted function with arguments its precondition
forbids raises a precondition failure, reported against the caller
(\cref{fig:contract-calls}). A postcondition thus binds only the function it
annotates, whereas a precondition additionally imposes an obligation on every
caller, an asymmetry we rely on in \cref{subsec:spec-synthesis}.

\begin{figure}[t]
\begin{lstlisting}[style=py]
def callee(x: int) -> int:
  '''pre: x >= 0'''
  return x

def caller(z: int) -> int:
  '''post: True'''
  return callee(z - 10)
# PreconditionFailed: "x >= 0" not satisfied
\end{lstlisting}
\caption{How CrossHair handles a callee's contract during a caller's
analysis. A callee's \emph{precondition} is checked at the call
site: reaching \texttt{callee} with a negative argument makes \texttt{caller}
fail, so an over-strong precondition shows up when a caller is analyzed.}
\label{fig:contract-calls}
\end{figure}

Because CrossHair's symbolic types
mirror the semantics of
Python's built-ins, any code whose behavior reduces to
those built-ins can be executed symbolically, including pure-Python
implementations drawn from the standard library and third-party
packages~\cite{schanely2017crosshair}. Conversely, code that reduces to Python outside of the types modeled by CrossHair cannot be
analyzed. The framework introduced in the remainder of this paper is aimed squarely at
recovering analyzability across this boundary.

\subsection{Target codebases}
\label{sec:bg-targets}
This work targets open-source Python web applications that interact with
databases. Such codebases are ill-suited to off-the-shelf symbolic analysis:
although they are widely used and actively maintained, their tight coupling to
stateful frameworks and database backends places much of their logic beyond the
reach of an engine that can only reason about self-contained code. Most of this
coupling runs through an \emph{object-relational mapper} (ORM): a library that
maps database rows to ordinary Python objects. To make this
concrete, consider the function in Fig.~\ref{fig:pure-vs-orm}, annotated with a CrossHair contract.
\begin{figure}[t]
\begin{lstlisting}[style=py]
def order_balance(order_id: int) -> int:
    '''
    pre: order_id > 0
    post: _ >= 0
    '''
    order = Order.objects.get(pk=order_id)
    return order.subtotal - order.discount
\end{lstlisting}
\caption{Contract-annotated function from a hypothetical store
application. \texttt{order\_balance} is coupled to the ORM and
cannot be executed without a model of the database environment.}
\label{fig:pure-vs-orm}
\end{figure}
CrossHair cannot execute this function symbolically. The call
\texttt{Order.objects.get(pk=order\_id)} dispatches into ORM
machinery, which expects a configured database connection and issues SQL against
a populated table. CrossHair has no built-in functionality to model the surrounding environment or database, so this will fail to execute. Notably, the post-condition is
exactly the kind of property we would like to check (a discount exceeding
the subtotal would make the balance negative), but it is unreachable without a
model of the ORM function.

\subsection{Coding agents and skills}
\label{subsec:bg-agents}
\chgi{A \emph{coding agent} is an LLM embedded in a control loop, or
\emph{harness}, that gives it tools to read and write files, run shell commands
and observe the results, so that it can carry out multi-step software
engineering tasks through an agent--computer interface~\cite{sweagent}.
\emph{Agent Skills} package the procedural knowledge for a specialized task into
a loadable unit~\cite{agentskills}: a directory with a \texttt{SKILL.md} of
natural-language instructions that the agent loads on demand, together with
optional scripts and references.}

\section{Methodology}
\label{sec:methodology}

\subsection{System Overview}
\label{subsec:sys}
\begin{chg}
\Cref{fig:system-overview} and \cref{alg:crosshair} show our pipeline in five
stages, which the rest of this section follows in order.
(1)~\emph{Environment synthesis} models the database and the framework so that
CrossHair can execute the code that uses them (\cref{subsec:env-modeling}).
(2)~\emph{Contract generation} writes a contract into each function under test,
guided by the schema from stage~1, and records functions the agent already
judges defective (\cref{subsec:spec-synthesis}).
(3)~\emph{Symbolic execution} runs CrossHair on every contracted function,
without model calls (\cref{subsec:symex}).
(4)~An \emph{audit} agent judges each CrossHair finding and each flagged defect,
and (5)~a \emph{witness} test confirms each real bug against the unmodified
application (\cref{subsec:auditing}). Each subsection names the steps of its
stage in \cref{alg:crosshair}, agent steps in \textsc{small caps} and script
steps in \textsf{sans-serif}. \Cref{subsec:harness} describes the harness that
runs them.

The stages follow one design principle: \emph{the LLM proposes, the solver
checks, concrete execution confirms}. Stages 1 and 2 propose an environment
model and a specification, either of which may be wrong. Stage~3 searches for
an input that violates the specification, and its verdict is only as good as
both. Stages 4 and 5 remove counterexamples caused by a miscalibrated contract
or by the over-approximate environment, the last by running each bug against
the real application.
\end{chg}

\begin{algorithm}[t]
  \ifshowchanges\color{blue}\fi
  \caption{The \emph{crosshair-bugs} pipeline, in the five stages of
  \cref{fig:system-overview}. \textsc{Small caps} are agent steps (a fixed
  prompt whose result must validate against a JSON Schema), \textsf{sans-serif}
  are script steps, and $\parallel$ marks one agent per item.
  $\textsf{Gate}_k(x)$ checks an agent-written artifact $x$ with a script
  (compilation, contract lint, CrossHair parse) and on failure hands the report
  to a repair agent, at most $k$ times. $\delta$ is the detected ORM,
  $\mathcal{F}$ its model files, $e$ the runtime configuration, $\Sigma$ the
  schema constraints, $S$ the stub module, $L$ the lazy-initialization plugin,
  $\beta$ the budget of functions to contract, $\mathcal{I}$ the defects the
  planner flags, and $\Omega$ and $\mathcal{B}$ CrossHair's outcomes and
  findings.}
  \label{alg:crosshair}
  \KwIn{codebase $C$}
  \KwOut{report $\mathcal{R}$, contracted source}
  \BlankLine
  \tcp{Stage 1: environment synthesis}
  $(\delta,\mathcal{F}) \leftarrow \textsf{DetectOrm}(C)$;\quad $e \leftarrow \textsc{Preflight}(C,\delta)$\;
  $\Sigma \leftarrow \bigcup\,\{\textsc{Schema}(m,\delta) \mid m \in \mathcal{F}\}$ \tcp*{$\parallel$}
  $S \leftarrow \textsf{Gate}_k\bigl(\textsc{Constrain}(\textsf{Gate}_k(\textsc{Stubs}(\mathcal{F},\delta)),\ \Sigma)\bigr)$\;
  $L \leftarrow \textsf{LazyInit}\bigl(\Sigma,\ \textsc{Invariants}(\mathcal{F},\Sigma)\bigr)$\;
  \BlankLine
  \tcp{Stage 2: contract generation}
  $\mathcal{T} \leftarrow \textsf{Rank}(\textsc{Targets}(C),\ \beta)$;\quad $\mathcal{I} \leftarrow \varnothing$\;
  \Repeat{$\textsf{\upshape Pending}(\mathcal{T}) = \varnothing$}{
    \ForEach(\tcp*[f]{$\parallel$}){$a \in \textsf{\upshape Chunk}(\textsf{\upshape Pending}(\mathcal{T}))$}{
      $(p_a, \mathcal{I}_a) \leftarrow \textsc{Plan}(a,\delta,\Sigma)$;\quad $\mathcal{I} \leftarrow \mathcal{I} \cup \mathcal{I}_a$\;
    }
    \lForEach{$g \in \textsf{\upshape ByFile}(\{p_a\})$}{$\textsf{Gate}_k(\textsc{Apply}(g))$\hfill$\parallel$}
    $\textsf{Gate}_k(\textsf{Lint}(C))$\;
  }
  \BlankLine
  \tcp{Stage 3: symbolic execution}
  $O \leftarrow \textsf{CrossHair}(C, S, L, e)$;\quad $(\Omega, \mathcal{B}) \leftarrow \textsf{Classify}(O)$\;
  \BlankLine
  \tcp{Stage 4: audit}
  $\mathcal{J} \leftarrow \textsc{Reconcile}\bigl(\{\textsc{Audit}(b) \mid b \in \mathcal{B} \cup \mathcal{I}\}\bigr)$ \tcp*{$\parallel$}
  \BlankLine
  \tcp{Stage 5: witness validation}
  \ForEach(\tcp*[f]{$\parallel$}){$b \in \mathcal{J}$ judged a real bug}{
    $w_b \leftarrow \textsc{WitnessUnit}(b)$\;
    \lIf{$w_b$ reproduces}{$w_b \leftarrow w_b \cup \textsc{WitnessE2E}(b)$}
  }
  $\mathcal{R} \leftarrow \textsf{Report}(\Omega, \mathcal{J}, w)$\;
  \Return $\mathcal{R}$, contracted source\;
\end{algorithm}

\subsection{Stage 1: Environment Synthesis (Alg.~1, lines~1--4)}
\label{subsec:env-modeling}

\begin{chg}
CrossHair can analyze only a function it can execute, and a function embedded
in a web framework cannot run until the framework is started and its database
modeled, work that has so far been written by hand. Our agent seeds this model
with prior knowledge: the skill packages what we know about each supported
framework (stub templates for ORM calls, start-up patterns, constructors for
framework objects). A script identifies the ORM and its model files
(\textsf{DetectOrm}), and an agent brings up the framework's runtime from the
repository's settings (\textsc{Preflight}). Agents then read the model
definitions and migrations for the database's constraints (\textsc{Schema}) and
encode them in a constrained stub module (\textsc{Stubs}, \textsc{Constrain})
and a lazy-initialization plugin (\textsf{LazyInit}). Every artifact an agent
writes passes a script check (compilation, contract lint, a CrossHair parse),
and the agent repairs reported failures.

The central obstacle is the database. A call into the persistence layer leaves
Python entirely, issuing SQL against a database that has no concept of symbolic
values. The same is true of any boundary the language cannot follow (an HTTP
request, a C extension, a read from the filesystem), and the technique we
describe could apply to all of them, but we model the database.
\end{chg}

Our approach, sketched in \cref{fig:system-overview}, is to have the agent interpose import-time \emph{stubs}, a form of mock installed by monkey-patching, on the ORM calls that the symbolic engine cannot execute. A stub intercepts the call that would have reached the database and supplies a value in its place, so that execution continues through the logic that consumes it. The decisive question is what that value should be.

A stub could return a \emph{concrete} value.
This lets the function run, but it collapses the symbolic state onto a single
path. At the opposite extreme, a stub could return an
\emph{unconstrained symbolic} value, which the engine is free to explore in
full. We take the middle path: the stub returns a \emph{constrained symbolic} value.
The agent searches the repository for evidence of constraints the database itself would enforce on any persisted row and imposes them on the symbolic value the stub returns.

\begin{chg}
\paragraph{Lazy initialization of ORM and framework objects}
Stubs supply values for calls \emph{into} the database, but many functions also
receive database-backed objects as arguments: a model instance as
\texttt{self}, a related object, a request or a form. CrossHair synthesizes
arguments from type annotations by running the class constructor on symbolic
inputs, which for ORM models and framework classes usually fails, so the
function is never analyzed. We therefore generate a plugin that applies lazy
initialization~\cite{khurshid2003generalized,deng2012efficient} through
CrossHair's \texttt{register\_type} hook, without modifying the engine. A model
instance is created without running its constructor. Each field is
materialized as a symbolic value on first read, and each related object is
built on first traversal, up to three levels deep. Values are bounded by the
harvested schema constraints and by per-field invariants that an agent extracts
from validators, choices and forms (\textsc{Invariants}), a form of bounded lazy
initialization~\cite{rosner2013bounded,rosner2015bliss,braione2013enhancing}.
In classic bounded lazy initialization these bounds, like the class invariants
of generalized symbolic execution, are written by hand, whereas here the agent infers
them from the code.
Classes whose constructor cannot run receive an under-constrained instance
whose attributes are synthesized on first access~\cite{ramos2015under}, and
framework objects (requests, sessions, forms, administrative contexts) receive
symbolic stand-ins. Stubbed query results are drawn from the same lazy
instances, so a lookup returns rows that satisfy it. The plugin is a fixed part of
the harness, the same for every project. Only the list of model classes and the agent-extracted invariants it is filled with are project-specific.
\end{chg}

This model is deliberately an over-approximation---meaning we constrain each stubbed call on its own, to the invariants the schema guarantees, \chgi{and beyond keeping repeated queries on one execution path consistent, we do not model the database as a stateful store.}
We accept instead that the over-approximation admits some infeasible executions and discharge the resulting false
positives downstream, through triage and concrete witness validation, rather
than through heavier stub machinery.

\subsection{Stage 2: Contract Generation (Alg.~1, lines~5--11)}
\label{subsec:spec-synthesis}

\chgi{CrossHair reports a violation only relative to a specification supplied from
outside~\cite{barr2014oracle}, and real-world code ships without one.} This stage
removes that barrier by having the agent synthesize the specifications. Given an existing function, the agent reads its body along with the surrounding context and emits a contract, annotating the source code without changing the code itself.  Concretely the agent enumerates the functions worth contracting (\textsc{Targets}), and a script ranks them, keeps at most $\beta$ (\textsf{Rank}) and divides them into batches (\textsf{Chunk}). Within a batch, a planning subagent drafts a contract for each function, using the schema constraints from stage~1 (\textsc{Plan}), and the resulting plans are grouped by source file (\textsf{ByFile}) so a single applier writes all of a file's contracts in one pass (\textsc{Apply}). \chgi{While planning, the agent also records functions it judges defective ($\mathcal{I}$), which are audited alongside CrossHair's findings.}

 A contract is valuable only insofar as it captures what a function is \emph{meant} to do, not what its code literally does. Recovering that intent is the part of the task we ask the agent to perform: it reads past the implementation, grounding the inference in type annotations, docstrings, parameter and return names, how callers consume the result, and the validation the code already performs, and states the abstraction behind it as a checkable property. This places real weight on the agent's ability to abstract, which varies from function to function.

\label{par:contract-calibration}
 Synthesis with an LLM is heuristic.
 A miscalibrated contract therefore surfaces as evidence to be triaged, not a verdict to be trusted. \chgi{Three of the four miscalibration modes are self-revealing.}
 \chgi{An over-permissive precondition or an over-strong postcondition yields a concrete counterexample, which the audit attributes to the contract rather than to the code, and} an over-restrictive precondition is exposed when an analyzed caller passes inputs that the precondition excludes, which CrossHair reports as a precondition failure in the caller (\cref{subsec:bg-crosshair}). The exception, however, is a postcondition that asserts too little: it raises no alarm and is a silent false negative.

\subsection{Stage 3: Symbolic Execution (Alg.~1, line~12)}
\label{subsec:symex}

\begin{chg}
With the instrumented repository in hand, the harness runs CrossHair on each
contracted function as a script step with no model calls (\textsf{CrossHair}).
Every contracted file has already passed the contract lint and a CrossHair
parse check (\textsf{Lint}). For each condition, CrossHair either finds a
\emph{counterexample}, a concrete input (including an initial database state)
that violates it or raises an exception the contract does not allow, or keeps
exploring without a violation until the search space or the budget runs out. A
search that exhausts its bounded state space is strong evidence, but not a
proof, that the function meets its contract under the harvested constraints.
\Cref{subsec:rq2} turns these outcomes into the exploration tiers we report.

A counterexample need not be a bug. It may come from a contract that is too
strong, from a stub that misses an ORM call or returns a row the real database
could not produce, or from a symbolic value that leaks into code the engine
does not model. Separating these from application bugs is the job of the
audit.
\end{chg}

\subsection{Stages 4--5: Audit and Witness Validation (Alg.~1, \mbox{lines~13--18})}
\label{subsec:auditing}
\begin{chg}
\paragraph{Audit (line~13)}
A script first sorts CrossHair's output into outcome classes and extracts the
findings, i.e.\ counterexamples and application crashes (\textsf{Classify}).
Each finding, together with each defect the planning agent recorded during
contract generation, is then judged by its own audit agent (\textsc{Audit}).
The auditor reads the code, the contract and the trace, and decides whether the
failure lies in the application, in the synthesized environment, or in the
generated contract. For a real bug it also assigns a severity on a four-level
scale given in its prompt (critical: data loss, a security issue or a silent
incorrect write; high: a wrong result or inconsistent state that propagates;
medium: observable wrong behavior; low: real but covered by upstream
validation), predicts whether the bug is reachable from a public entry point or
latent, and records whether its discovery depended on the ORM model. Both
labels are the LLM's judgment, and no human reviews them during the run. A consistency pass then reviews the verdicts of each
package together and corrects contradictions (\textsc{Reconcile}). Findings
attributed to the environment or to the contract are not reported.

\paragraph{Witness validation (lines~14--18)}
Each bug the audit judges real then receives an executable test against a
non-symbolic instance of the application. A unit witness calls the function
directly (\textsc{WitnessUnit}). Every bug it reproduces also receives an
end-to-end witness that tries to trigger the failure through a public entry
point of the running application (\textsc{WitnessE2E}), which decides whether
the bug is user-facing or latent and checks the audit's prediction. A bug
counts as confirmed only if a witness reproduces it, which grounds every
reported finding in concrete program behavior rather than in an artifact of the
symbolic environment.

\subsection{Harness Implementation}
\label{subsec:harness}

We implement \cref{alg:crosshair} as an \emph{agent harness}: the program
around the model that decides which model calls are made, what each one
receives, and what is done with its output~\cite{lopopolo2026harness}.
Following~\cite{schluntz2024agents}, we build the pipeline as a
\emph{workflow}, in which LLM calls are orchestrated through predefined code
paths, as Agentless does for issue repair~\cite{xia2025agentless}, rather than
letting an agent direct its own process. This trades flexibility for the
predictability a measurement study requires. All state lives on disk as
structured JSON~\cite{young2025harnesses}, so an interrupted run resumes
without repeating finished work. Every count we report is computed by a script,
not by the model, and the agent that judges a result is never the agent that
produced it~\cite{rajasekaran2026harnessdesign}. Because the control flow lives
in code, two runs on the same repository walk the same graph over the same work
list and differ only in the content of individual model calls. Each run records
the hashes of its workflow scripts, the tool versions, and per-phase token and
time costs. The graphs are JavaScript programs for Claude Code's Workflow tool
and use no third-party orchestration framework.
\end{chg}

\subsection{Baseline: the ai-bugs skill}
\label{subsec:ai-bugs}
For comparison we evaluate \texttt{ai-bugs}, a purely LLM baseline that uses no
contracts and no symbolic execution (\cref{alg:aibugs}). It discovers the
source files (\textsc{Discover}), fans four bug-class subagents across each file
(logic errors, edge cases, data integrity, error handling; \textsc{Find}),
filters their findings with a skeptical verifier (\textsc{Screen}), and compiles
the survivors into a report (\textsc{Compile}). \chgi{It uses the same model and
the same 1,429 files our pipeline contracted, in separate sessions with no access
to our results. Every baseline bug in a function our pipeline did not report is
witnessed with the same unit and end-to-end procedure as ours.}

\begin{algorithm}[t]
  \caption{AI multi-agent bug finding, performed with no symbolic execution}
  \label{alg:aibugs}
  \KwIn{codebase $C$}
  \KwOut{screened bug report $\mathcal{R}$ of candidate bugs}
  
  $\mathcal{F} \leftarrow \textsc{Discover}(C)$;\quad $\mathcal{S} \leftarrow \varnothing$\;
  \ForEach{$f \in \mathcal{F}$}{
    $\mathcal{B}_f \leftarrow \bigcup_{\kappa \in K} \textsc{Find}(f, \kappa)$\;
    \lIf{$\mathcal{B}_f \neq \varnothing$}{$\mathcal{S} \leftarrow \mathcal{S} \cup \textsc{Screen}(f, \mathcal{B}_f)$}
  }
  $\mathcal{R} \leftarrow \textsc{Compile}(\mathcal{S})$\;
  \Return $\mathcal{R}$\;
\end{algorithm}

\section{Experimental Evaluation}
\label{sec:experiments}

We answer the three research questions of \cref{sec:introduction}. The unit
of analysis throughout is a \emph{function}, with each function having a single
contract.

\subsection{Setup and Reproducibility}
\label{subsec:setup}

\begin{chg}
All experiments use \texttt{Claude Opus 5.5}, run through
\texttt{Claude Code} version \texttt{2.1.x}. %
Symbolic execution is performed by CrossHair version \texttt{v0.0.104}, dispatching path constraints to Z3 version \texttt{4.16.0}, with a budget of 30 seconds per condition and a per-function cap of 300 seconds, raised up to 900 seconds for functions with several postconditions.
Each repository is a fresh clone at the study commit, with a virtual environment
rebuilt from the original package freeze.
The stubs, plugin and contracts under evaluation are those the harness
produced, with no manual edits to its output.
Within a run there is no open-ended triage loop: every repair loop is bounded
($k=2$ attempts per gate), and contract generation is relaunched until no work
is pending. The contract budget is 800 functions per repository, ranked by a
static risk score. For zulip, the largest application, we ran a second batch of
800, giving \numContracted\ contracted functions in total.
Bugs are counted per function, as the union over a
repository's runs.
The pipeline's LLM cost, counted per harness run, was at least \$2{,}821 for
all eleven repositories (a lower bound, as it excludes the orchestrating session). The baseline (\cref{subsec:ai-bugs}) cost about \$712, plus \$925 to witness its
findings. These costs are estimated from the per-token API prices at the time of the runs, although all runs were made under a Claude subscription rather than paid
per token.
\end{chg}

To enable independent replication, we provide a supplementary
artifact\footnote{\url{https://doi.org/10.6084/m9.figshare.32835236}} containing the orchestration
skills, compact captures of all instrumented repositories (as patches pinned to
upstream commits), per-function run data, the AI-only baseline with its
witnesses, and the scripts that compute every reported number.

\subsection{RQ1: Reach}
\label{subsec:rq1}

\begin{chg}
\paragraph{Terminology}
A contracted function is \emph{executed} if CrossHair completes at least one
path through it on which a postcondition is checked, or refutes one of its
conditions. It is \emph{blocked} otherwise: the module fails to import or the
framework fails to start, no input satisfying the precondition can be built,
or every path aborts before the body returns.

\paragraph{Environment ablation}
\label{par:reach-ablation}
We run every contracted function under three configurations, with the same
contracts, budget and classifier.
\emph{Bare} runs plain \texttt{crosshair check}, with no plugins.
\emph{Setup} adds only framework initialization, done the way the project's own
test suite does it (settings module and \texttt{django.setup()}, or the
application factory).
\emph{Full} is \emph{crosshair-bugs}: the start-up plugin, the
schema-constrained ORM stubs, lazy initialization and framework stand-ins. The setup arm separates the generic part of the
environment, which a developer would write once per project, from what the
agent synthesizes.
\end{chg}

\subsection{RQ2: Exploration Depth and Outcomes}
\label{subsec:rq2}
\begin{chg}
We assign each contracted function one
\emph{exploration tier}, evaluated on the contract set the harness produced
(\cref{subsec:harness}). Miscalibrated contracts are not corrected within the
run but are identified by the audit (\cref{par:contract-calibration}).

\paragraph{Exploration tiers}
Let $c$ be the number of distinct paths on which a postcondition was checked
and held, read from CrossHair's path-tree statistics, and taken as the minimum
over the function's postconditions.
\begin{itemize}
\item \textbf{Exhausted:} every path was explored and every condition held,
      relative to the environment model.
\item \textbf{Bounded:} $c \ge 25$ and no condition refuted.
\item \textbf{Shallow:} $1 \le c < 25$ and no condition refuted.
\item \textbf{Refuted:} CrossHair reported a counterexample, an uncaught
      exception or a violated callee precondition. These are the \emph{candidate findings}, which the audit splits into real bugs,
      false positives, tool artifacts (filtered automatically) and failures
      located only inside an installed library.
\item \textbf{Blocked:} $c = 0$ and nothing refuted (\cref{subsec:rq1}).
\end{itemize}
The thresholds count paths, not a share of them: unless the search is
exhausted, the total number of paths is unknown. A function with few paths can
therefore be shallow even if the paths checked are most of them, for example 9
of 10 when the search stops before the last. A function whose search finishes
is exhausted whatever its $c$.
\end{chg}

\subsection{RQ3: Bug Yield}
\label{subsec:rq3}
A bug is \emph{confirmed} when the audit judges it real (\cref{subsec:auditing})
and a witness reproduces it against the real application code. A confirmed bug
is \emph{user-facing} if its end-to-end witness triggers it through a public
entry point, and \emph{latent} otherwise. We report the confirmed bugs (RQ3.1),
describe them (RQ3.2), and compare them with those of the inspection-only
baseline (\cref{subsec:ai-bugs}) (RQ3.3).

\section{Results}
\label{sec:results}

\subsection{RQ1: Reach}
\label{sec:results-rq1}
\begin{chg}
The synthesized environment is what lets CrossHair run on these applications.
Of the \numContracted\ contracted functions, bare CrossHair
executes 243 (\pctExecutedLite\%), framework initialization alone 1{,}595
(\pctExecutedSetup\%), and the full synthesized environment 4{,}394
(\pctExecuted\%) (\cref{fig:rq1-reach}).

  \begin{figure}[t]
    \centering
    \includegraphics[width=0.82\columnwidth]{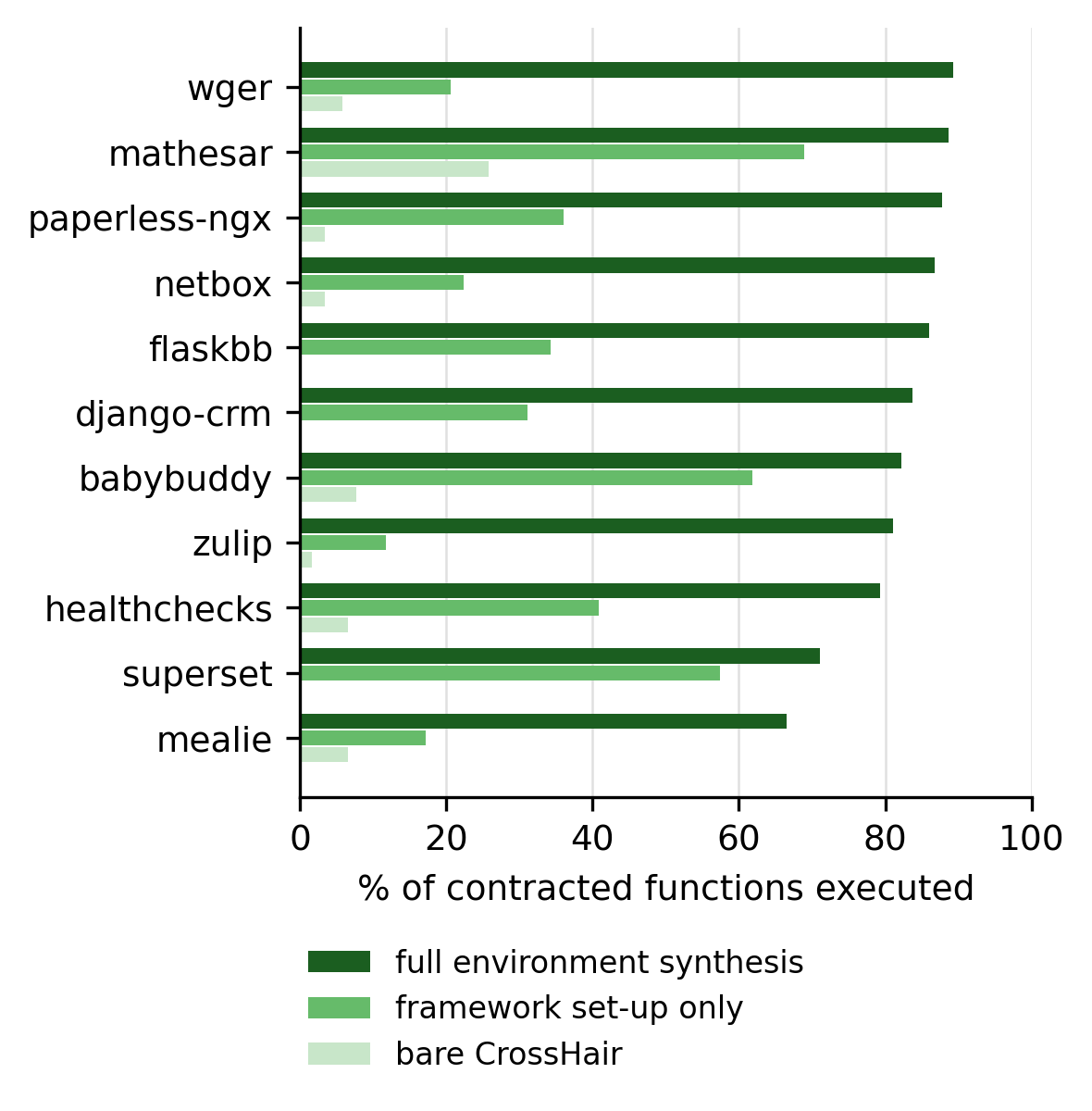}
     \caption{Executed contracted functions per repository under three
     environments: bare CrossHair, framework initialization only, and full
     environment synthesis (\emph{crosshair-bugs}). Pooled over
     \numContracted\ functions: \pctExecutedLite\%, \pctExecutedSetup\% and
     \pctExecuted\%.}
    \label{fig:rq1-reach}
  \end{figure}

\paragraph{Bare and framework set-up}
Bare CrossHair executes almost nothing: in three repositories not a single
function, because the modules cannot be imported without a configured
framework. What it does execute is code that never touches the framework, such
as helpers that build SQL expressions without an application context.
Framework initialization, the part a developer would
write once per project, recovers about a third of the functions. What remains
blocked under it is mostly argument construction: model instances whose
constructor needs framework state, and string-typed annotations that CrossHair
cannot resolve.

\paragraph{Full environment synthesis}
These are exactly what the agent-synthesized environment supplies: lazy
initialization of model instances, ORM stubs and framework stand-ins. With them, every repository executes between 66\%
and 89\% of its functions. The gain over framework set-up is largest where the
code is most object-heavy, up to 69 percentage points (from 12\% to 81\%).
\end{chg}

\subsection{RQ2: Exploration Depth and Outcomes}
\label{sec:results-rq2}
\begin{chg}
\Cref{fig:rq2} gives the tier of every contracted function. Pooled, 39.1\%
are refuted, 28.6\% bounded, 10.0\% shallow and 3.6\% exhausted; 18.7\% remain
blocked.

  \begin{figure}[t]
    \centering
    \includegraphics[width=0.82\columnwidth]{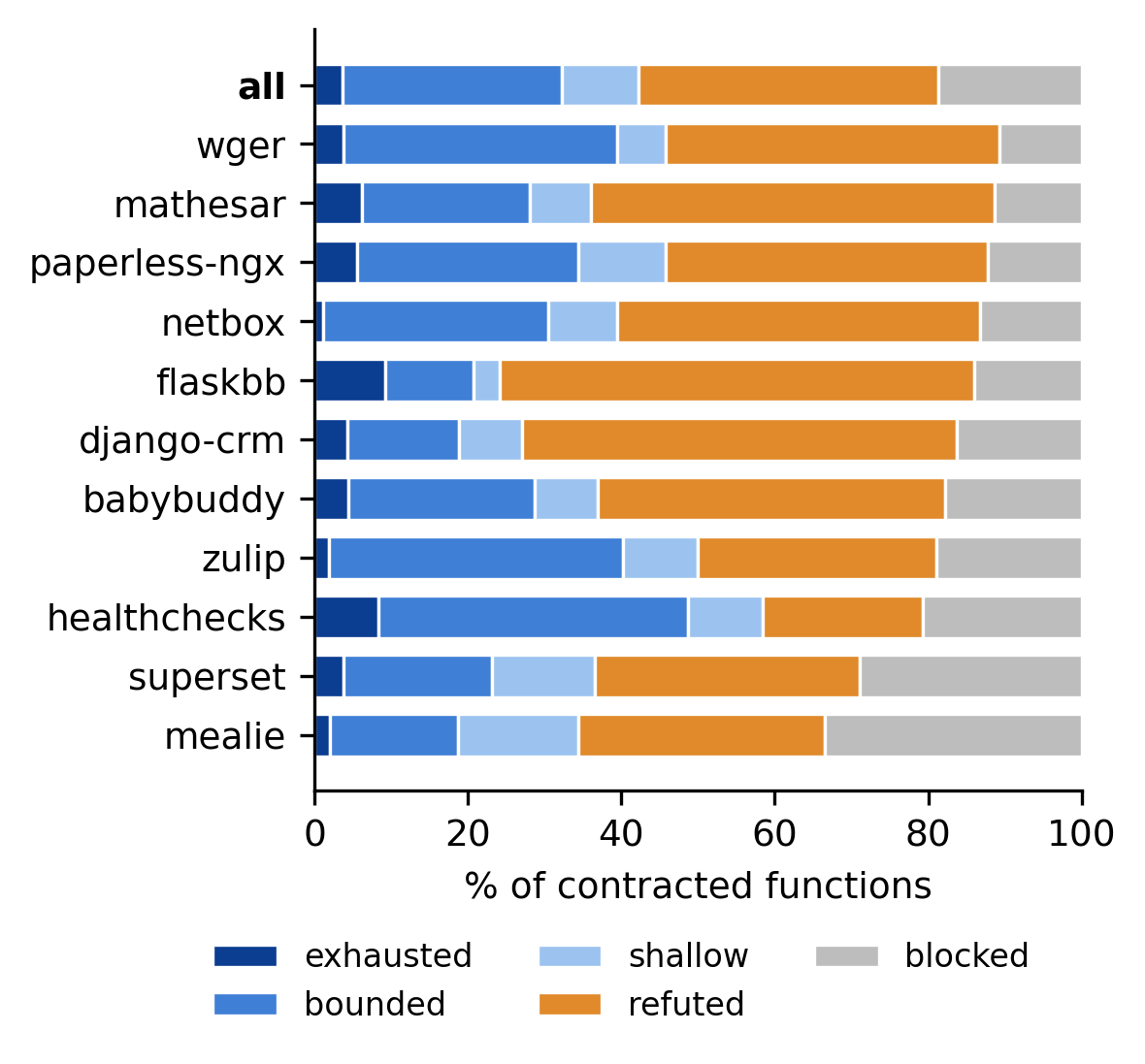}
    \caption{Exploration tier of each contracted function, per repository and
    pooled ($n{=}5{,}406$). Exhausted: all paths explored; bounded: at least 25
    checked paths; shallow: 1--24; refuted: a condition was violated
    (candidate finding); blocked: no checked path.}
    \label{fig:rq2}
  \end{figure}

\paragraph{Refutations are mostly not bugs}
The audit splits the 2{,}113 refuted functions into 241 real bugs (4.5\% of all
functions), 1{,}281 false positives (23.7\%), 562 tool artifacts filtered
without an audit (10.4\%) and 29 failures located only inside an installed
library (0.5\%). False positives are the price of an over-approximate
environment: a lazily built instance or a stubbed query result that the real
database could not produce, a symbolic value escaping into code that expects a
concrete one, or an input that the application's own validation rejects
earlier. This is why every CrossHair finding is audited, and every finding the
audit accepts as a bug is then witnessed.

\paragraph{Exploration is bounded, not exhaustive}
Only 3.6\% of functions are explored exhaustively; for most executed functions
CrossHair checks many paths (the per-repository median ranges from 27 to 726) but does
not finish within budget. A bounded verdict is therefore evidence, not a proof.
The obstacles here are the classic ones of symbolic execution: path explosion,
and operations outside the solver's theories (string and regular-expression
manipulation, date parsing, hashing, C extensions), which CrossHair concretizes. \chgi{Each maps onto a mature line of symbolic-execution
research~\cite{baldoni2018survey}, such as state
merging~\cite{kuznetsov2012efficient,trabish2023state} and
abstraction~\cite{anand2009symbolic,shannon2007abstracting}, which an agent
could in future instantiate in the same way as lazy initialization.}

\paragraph{Blocked functions}
The remaining 18.7\% are blocked mostly by gaps in the environment model or by
contract defects, not by limits of symbolic execution. Further repair passes
on the environment and contracts could recover many of them, but we chose not
to spend the compute.
\end{chg}

\subsection{RQ3: Bug Yield}
\label{sec:results-rq3}
\begin{figure}[t]
\centering
\includegraphics[width=0.82\columnwidth]{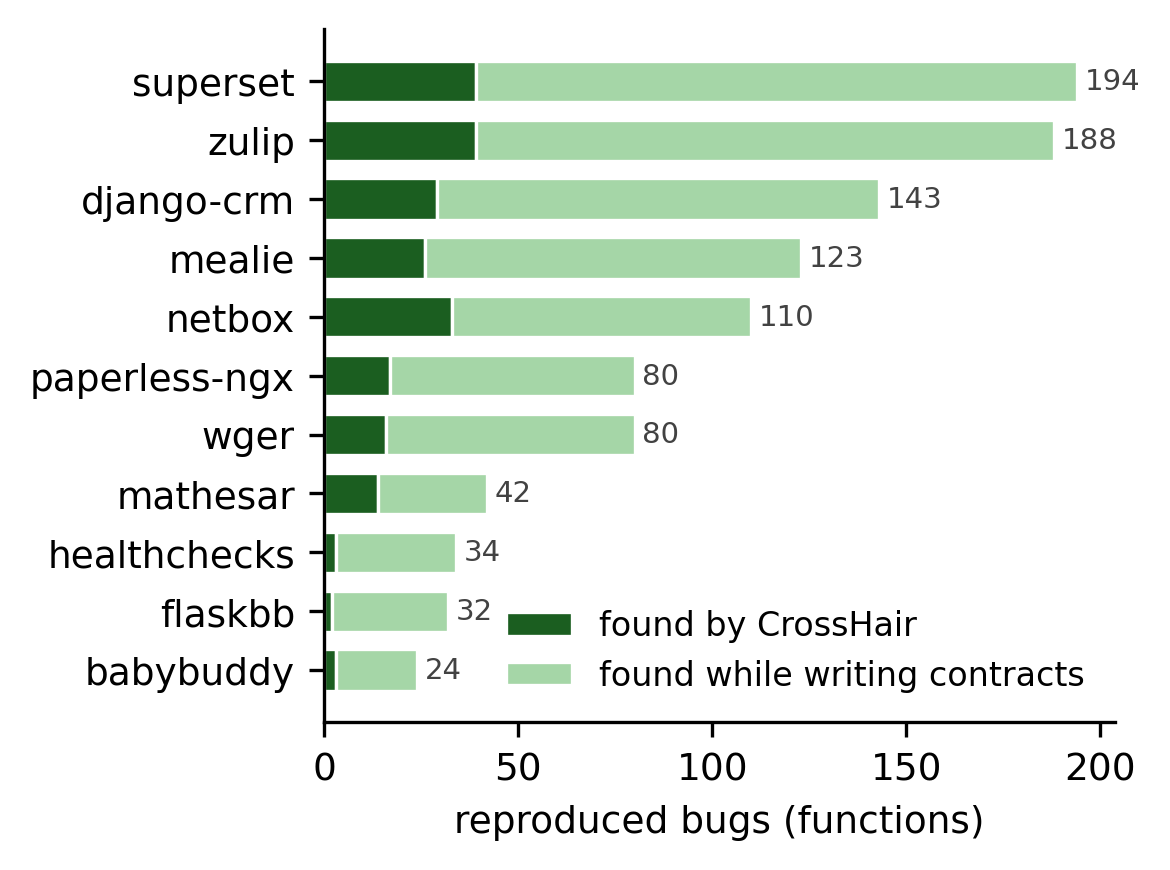}
\caption{\chgi{Reproduced bugs per repository by discovery mechanism: found
by CrossHair (a counterexample or uncaught exception), or flagged by the agent
while writing contracts.}}
\label{fig:rq3-discovery}
\end{figure}

\begin{chg}
The pipeline found \textbf{\numReproducedBugs\ witness-reproduced bugs} across
the 11 repositories: \numCHBugs\ found by CrossHair and \numSpecBugs\ flagged
by the agent while writing contracts. Our interest is mainly in the former,
the bugs symbolic execution finds, but we report both because both pass the same
audit and witnesses.

\paragraph{From candidates to confirmed bugs (RQ3.1)}
Two sources feed the audit. Here a bug is counted at the function that
contains it, which may be a callee of the function CrossHair was analyzing (RQ2 counts the analyzed function instead). After automatic filtering of tool
artifacts, CrossHair's findings point to 1{,}609 functions, of which the audit
judged 226 to contain a real bug (14\%). The planners flagged 1{,}013 functions while writing
contracts, of which the audit accepted 991 (98\%). Together, 1{,}064 functions
were judged to contain a real bug.

\paragraph{Witness validation}
For each of these the agent wrote a unit test that runs the real code, with no
stubs. \textbf{\numReproducedBugs\ were reproduced; 14 were not}. We treat the
\numReproducedBugs\ reproduced bugs as the confirmed yield. The agent then
drove each confirmed bug through a public entry point of the running
application with a real database: \textbf{\numEndToEndBugs\ (\pctEndToEnd\%)
reproduced end-to-end}. The audit's reachability labels largely held: of the
bugs it had labeled user-facing, 889 of 912 were confirmed end-to-end, and 94
of the 137 it had labeled latent could not be reached. Of 45 high-ranked
confirmed bugs that we checked against each project's current code,
maintainers had since fixed 22, with commits that describe the fix.

\paragraph{Blind human review}
One of the authors rated a stratified random sample of 50 of our confirmed bugs and
50 of the baseline's, shuffled and without method labels. They judged 72\% of ours and
70\% of the baseline's real, with no significant difference, but
rated 31 of ours as crashes against 12 of the baseline's, whose most common rating was
minor. Rejected bugs were mostly judged intended or unreachable in practice,
often because the user interface validates input the API admits. Across both
methods, bugs reproduced end-to-end were judged real 82\% of the time, those
reproduced only in a unit test 4 of 18.

\paragraph{Qualitative description (RQ3.2)}
Discovery is dominated by contract writing: \numSpecBugs\ of the bugs (79\%)
were flagged by the agent while it wrote contracts, and \numCHBugs\ (21\%) were
found by CrossHair, 69 of them in functions the planner had not flagged
(\cref{fig:rq3-discovery}). By severity, as assigned by the audit, 689 are
low, 321 medium, 39 high and 1 critical. The audit labels 204 bugs as coupled
to the ORM, 84 of them found by CrossHair.

The bugs CrossHair finds are \textbf{input-domain corner cases}: mostly
missing or \texttt{None} fields, empty or whitespace-only strings, unusual
characters (NUL, control or non-ASCII) and malformed formats. They
share a root cause: a value is accepted with the right \emph{type} but not
sufficiently validated, while downstream code assumes more
(\cref{fig:bug-examples}). An SMT solver
finds them because it explores values that validation admits but tests rarely
probe. Of the 212 CrossHair bugs whose function we could match across the
ablation arms, 110 (52\%) lie in functions that are executable only with
the synthesized environment: they are found only because the agent
synthesized it.
\end{chg}

  \begin{figure*}[t]
  \small
  \textbf{(a)}~superset \texttt{cast\_to\_num}: an unintended Unicode symbol slips past the type guard
  \begin{lstlisting}[style=py]
  if value.isdigit():     # True for '²'
      return int(value)   # int('²') -> ValueError  (HTTP 500 on /chart/data)
  \end{lstlisting}

  \textbf{(b)}~django-crm \texttt{Currency.save}: a copy-paste typo corrupts persisted data
  \begin{lstlisting}[style=py]
  # post: implies(__old__.self.is_state_currency,
  #               self.rate_to_state_currency == 1.0)   # -> counterexample
  if self.is_state_currency:
      self.is_state_currency = 1.0   # BUG: should be rate_to_state_currency
  \end{lstlisting}

  \chgi{\textbf{(c)}~zulip \texttt{check\_password\_strength}: found only by CrossHair}
  \begin{lstlisting}[style=py]
  # new_password has no length limit in the API
  zxcvbn(password, max_length=settings.PASSWORD_MAX_LENGTH)  # = 100
  # 108 chars -> ValueError (HTTP 500 on PATCH /api/v1/settings)
  \end{lstlisting}

  \caption{\chgi{Three} representative confirmed bugs. \textbf{(a)} Symbolic execution finds an input edge a type guard admits: \texttt{str.isdigit()} accepts the superscript
  digit \texttt{'²'} that \texttt{int()} rejects. \textbf{(b)} A synthesized
  postcondition exposes a typo that writes the rate into the boolean
  flag, silently corrupting persisted data. \chgi{\textbf{(c)} CrossHair generated a
  password longer than \texttt{zxcvbn}'s length limit, and neither the planner nor
  the inspection baseline flagged this function.}}
  \label{fig:bug-examples}
  \end{figure*}

\begin{figure}[t]
\centering
\includegraphics[width=0.78\columnwidth]{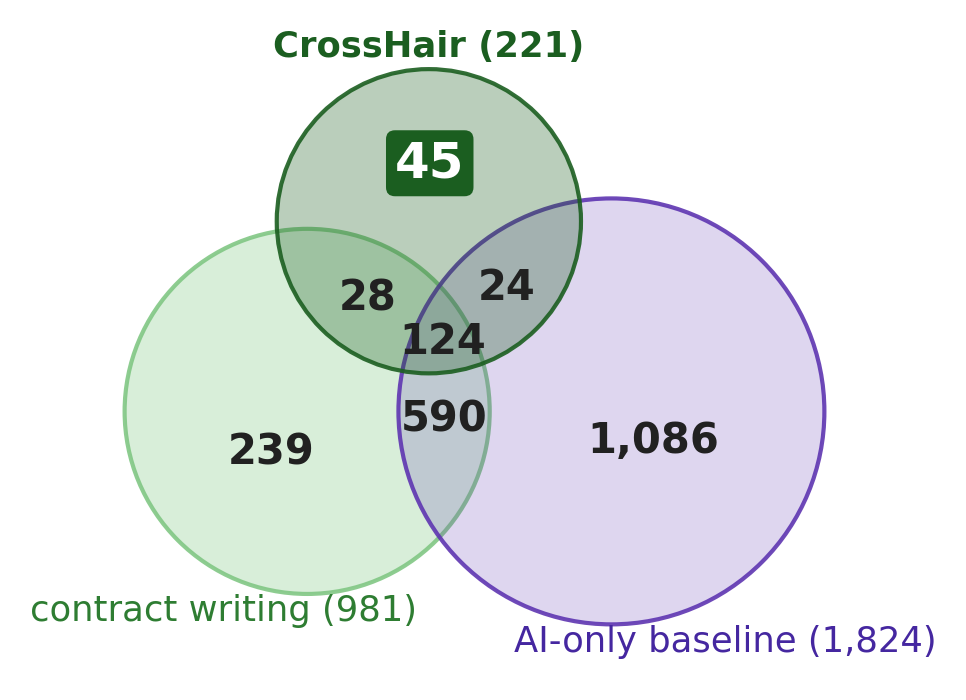}
\caption{\chgi{Reproduced bugs (functions) found by CrossHair, while writing
contracts, and by the AI-only baseline with the same model (areas not to scale).}}
\label{fig:rq3-overlap}
\end{figure}

\paragraph{Comparison against the baseline (RQ3.3)}
\begin{chg}
The baseline's verifier accepted 4{,}125 bug reports in 1{,}864 functions. Of
these, 738 are functions where our pipeline also reproduced a bug. We witnessed the remaining 1{,}126, and 1{,}086 (96\%) were reproduced, 925 of them
end-to-end. The baseline therefore reproduces \textbf{\numBaselineBugs} bugs,
more than our pipeline's \numReproducedBugs\ (\cref{fig:rq3-overlap}).
Our pipeline finds \textbf{\numAdditionalBugs\ bugs the baseline misses}, so
running both finds \pctImprovement\% more bugs than the baseline alone; 257 of
these are reproduced end-to-end.
Of the \numAdditionalBugs, 239 were flagged while writing contracts and 73 by
CrossHair. Relative to its size, CrossHair's contribution is the least
redundant: of the 69 bugs only CrossHair found, the baseline misses 45 (65\%),
against 29\% of the bugs found only while writing contracts
(\cref{fig:rq3-overlap}).

The overlap is measured per function, and the two bug sets differ in kind:
57\% of the baseline's witnesses assert a wrong output and 29\% a crash, while
ours are 56\% crashes and 30\% wrong outputs (\cref{sec:threats}). With a current frontier model,
direct inspection therefore finds more bugs, and more wrong-output bugs in particular, while symbolic execution over a synthesized environment remains
\textbf{complementary}: a third of its bugs are ones inspection misses.

\paragraph{Bugs only symbolic execution finds}
The strictest test of what symbolic execution adds is the set of bugs that
neither the planner, while writing contracts, nor the baseline flagged.
There are \textbf{45} such bugs. They
are not marginal: 37 are reproduced end-to-end, and the audit rates 8 medium
and 1 high severity. 26 of the 45 lie in
functions that are executable only with the synthesized environment, so they
depend on it as well as on the solver.
Almost all are input-domain corner cases that a reader of the function
overlooks because the offending input looks benign or out of scope. A
specification parser crashes on an empty list such as \texttt{\{"max": []\}},
a malformed URL such as \texttt{//[} makes \texttt{urlparse} raise, and a
password longer than a library's undocumented limit causes a server error
(\cref{fig:bug-examples}c). In each case the violating input passes every type
and validation check before the failing line.
\end{chg}

\section{Threats to Validity}
\label{sec:threats}

\begin{chg}
\paragraph{Construct validity}
Our oracle is synthesized, not given: each contract is the agent's
reconstruction of a function's intended behavior, so a miscalibrated contract
can manufacture a finding that no developer would call a bug. Witness
validation (\cref{subsec:auditing}) removes the findings that do not reproduce
against the unmodified application, but a reproduced violation can still be one
that no user can trigger, or behavior the developers intended: the blind
review judged 82\% of the sampled bugs that reproduced end-to-end to be real,
but only 4 of the 18 that reproduced only in a unit test
(\cref{sec:results-rq3}). The dual
error is not caught: a postcondition that asserts too little yields no
counterexample (\cref{par:contract-calibration}), so our counts are a
precision-oriented lower bound, not a recall measure. ``Executed'' (RQ1) is
only a proxy for analyzability, and RQ2 shows that executing a function is far
from exploring it exhaustively. Finally, no maintainer has reviewed our reports. The 22
upstream fixes (\cref{sec:results-rq3}) were made independently of this work,
and the blind review was done by one of the authors, who could not see which
method found each bug but knows both pipelines.

\paragraph{Internal validity}
Our RQ1 ablation separates framework initialization from the agent-synthesized
environment, but not the components of the latter (stubs, lazy initialization,
stand-ins) from one another, so the
$\pctExecutedSetup\%\rightarrow\pctExecuted\%$ gain is attributable only to them
jointly. The setup arm follows each project's test configuration, and a
developer might do better by hand. The split between bugs found by CrossHair
and by the agent while contracting reflects where each bug surfaced in one
pipeline, not an independent isolation of the two effects.

\paragraph{Conclusion validity}
LLM agents are stochastic, and we report one configuration per repository, with
bugs pooled over its runs (\cref{subsec:setup}). We do not report variance
across repeated runs because of their cost, which is the most direct threat to
replication. Cross-method comparisons other than the human review are
descriptive.

\paragraph{Baseline asymmetry and external validity}
The 738 functions in both bug sets were witnessed only through our pipeline,
and are counted for the baseline without checking that it reports the same
bug. Wrong-output witnesses, more common in the baseline, rest on the agent's
reading of intended behavior, which the human review checks only on a sample. Both
methods ran only on the files our pipeline contracted. Our corpus is eleven
popular open-source Python web applications on two ORMs (Django and
SQLAlchemy). We make no claim beyond ORM-backed Python, and the technique's
applicability to other boundaries (HTTP, C extensions, the filesystem) is not
evaluated. Popular projects may also be better typed and documented than
typical code, which aids both contract synthesis and CrossHair's type-driven
proxies. Finally, our figures are for a single frontier model in one harness,
at one point in time.
\end{chg}

\section{Related Work}
\label{sec:related_work}

\begin{chg}
\subsection{Specification Synthesis}
A growing body of work uses LLMs to synthesize the specifications that serve as
the \emph{oracle} for a downstream verification
tool~\cite{ma2025specgen,ugarefun2spec,suresh2026agentic,maaz2025agentic,granberry2026seeking}.
LLM-written postconditions can distinguish buggy from correct
code~\cite{endres2024nl2postcond}, and Richter and Wehrheim check such contracts
with CrossHair on small Python benchmarks~\cite{richter2025nl2contract}. We
apply the same kind of contract to large applications whose functions cannot
run without an environment model, and we synthesize that environment as well
as the oracle. LLM-generated oracles tend to encode the behavior the code has
rather than the behavior it should have~\cite{konstantinou2024oracles}, one
reason we audit every finding and require an executable witness.

\subsection{Bug-Finding Engines}
A synthesized oracle and environment could in principle drive any of several
bug-finding engines, such as symbolic or concolic execution, property-based
testing, fuzzing~\cite{park2026agentic, zhang2024fixing, jiang2024fuzzing}, or
deductive verification with a theorem prover~\cite{lucas2026formal}. The
closest agentic alternative is property-based testing with agent-written
properties: Maaz et al.~\cite{maaz2025agentic} have an agent infer properties of
Python library functions and check them with Hypothesis. We build on symbolic
execution for its natural fit with declarative contracts and its concrete
counterexamples. We do not compare it experimentally with property-based
testing or the other engines. Each is a mature methodology with a long history
of comparison, and our aim is to make symbolic execution run on code it
previously could not, not to argue that it finds better bugs than the
alternatives.

\subsection{Environment Modeling for Symbolic Execution}
Symbolic execution of database applications has a long history. Emmi et
al.~\cite{emmi2007dynamic} generate program inputs and database states
together for programs that issue SQL, Pan et al.~\cite{pan2013guided}
synthesize the database interactions that guide test generation, and
Marcozzi et al.~\cite{marcozzi2015relational} execute SQL code symbolically
over a relational model. SynthDB~\cite{chen2023synthdb} synthesizes database
contents by program analysis to test web applications. Beyond databases,
mocking frameworks such as Moles~\cite{dehalleux2010moles} isolate code from its
environment, and environment assumptions make unit proofs
tractable~\cite{priya2024vmocks}. For complex object inputs, generalized
symbolic execution~\cite{khurshid2003generalized,deng2012efficient}, bounded
lazy initialization~\cite{rosner2013bounded,rosner2015bliss,braione2013enhancing}
and under-constrained execution~\cite{ramos2015under} build inputs on demand.
In all of these, the model of the environment, or the invariants that bound it,
is an \emph{input} written by a human, and the database techniques target SQL
strings in Java or C\# programs. In our setting there is no query text: the
database is reached through ORM call sites whose framework semantics must be
modeled. We make the model an \emph{output}, synthesized by an agent from
framework knowledge and repository inspection, and we instantiate the
lazy-initialization techniques above with bounds the agent infers from the
code.

Recent work also uses LLMs to write the models a symbolic engine needs.
SAILOR~\cite{shafiuzzaman2026sailor} has an LLM synthesize and refine KLEE
harnesses (drivers, stubs and assertions) from engine feedback and confirms
each crash by concrete replay, finding far more memory-safety vulnerabilities
in C/C++ projects than an agentic baseline. Gordian~\cite{bouras2026gordian}
has an LLM write solver-friendly stand-ins for code the solver handles poorly
and partition unbounded heaps. Both target C programs and their memory-safety
or coverage goals, whereas we target the ORM and framework layer of Python web
applications.

\subsection{Testing Database-Backed Applications}
Database-backed web applications frequently declare data constraints
inconsistently across the database, the model and the user interface, and
handle their violation poorly~\cite{yang2020managing}, and our bugs cluster at
the same boundaries. Search-based system testing can treat the database as part
of the test input, inserting rows that steer
execution~\cite{arcuri2020handling}, but it runs a real database, where we use
an over-approximate symbolic model. Differential testing has been applied to
the ORM implementations themselves, including Django's and SQLAlchemy's, with
solver-generated records~\cite{sotiropoulos2021cynthia}, whereas we test the
applications built on them. For Python, automated unit-test generation has been
evaluated at scale~\cite{lukasczyk2023pynguin}, and symbolic path constraints
have been used to guide LLM test generation~\cite{ryan2024symprompt}, but
neither addresses code that cannot run without a modeled database and
framework.

\subsection{Agentic Path Exploration}
Other work places the LLM \emph{inside} the symbolic-execution loop to drive an
already-running engine deeper. Agentic concolic execution~\cite{luo2026agentic}
has agents summarize execution traces and synthesize new path constraints, and
Cottontail~\cite{tu2025cottontail} uses LLMs to shape highly structured inputs
that get past complex parsers. LLMs remain unreliable at the precise reasoning
full symbolic simulation needs~\cite{feng2025can}, so these systems use the LLM
as a heuristic guide. Others let the LLM take the solver's place and reason
about paths directly~\cite{li2025llmsymex}, or let it write specifications that
a classical analyzer then checks, as IRIS does with taint specifications for
CodeQL~\cite{li2025iris}. We place the agent \emph{around} a sound engine and
use it only to supply what the engine needs to run: where LLM-guided
exploration makes the engine explore better, we make it \emph{applicable}, and
an LLM error surfaces for us as a triageable false positive rather than an
unsound analysis step. The two are complementary, and a tool such as Cottontail
could exercise contracts whose inputs our engine cannot yet construct
(\cref{subsec:rq2}). Concurrent work applies the same division of labor to
bounded model checking of C and Rust code, adding a refinement loop over the
specifications~\cite{sun2026agenticmc}. It does not measure how much of the code
becomes analyzable or compare against LLM inspection with the same model, and
we do both. The same division underlies recent LLM-driven formal verification
at scale: an LLM wrote machine-checked Rocq proofs that 239 fuzz-test entry
points across 123 open-source libraries cannot crash, finding hundreds of bugs
along the way~\cite{lucas2026formal}.

\subsection{Neural Code Auditing and Agentic Vulnerability Discovery}
LLM agents with code navigation and testing tools can find vulnerabilities in
real codebases~\cite{repoAudit,glazunov2024bigsleep}, including zero-days found
and patched autonomously in the AI Cyber
Challenge~\cite{zhu2026teams,fang2024llm}. A key challenge is false positives,
which multi-agent cross-verification aims to
reduce~\cite{padmanabhuni2026meta}. LLMs are also used to triage the warnings of
static analyzers~\cite{li2024llift,li2025buglens,du2026reducing}, the role our
audit plays for CrossHair's findings, and to turn bug reports into executable
reproducing tests~\cite{mundler2024swtbench,zhao2026anypoc}, as our witnesses
do.
\end{chg}

\section{Conclusions}
\label{sec:conclusions}

\chgi{We showed that an agent can supply the environment models and contracts
that symbolic execution needs, in place of a human, at the scale of more than
five thousand functions. On eleven web applications, symbolic execution over the synthesized environment executes \pctExecuted\% of the contracted functions, against \pctExecutedLite\% without it, and CrossHair finds \numCHBugs\ witness-reproduced bugs. A baseline that asks the same model to find bugs directly reproduces more bugs overall (\numBaselineBugs), but misses a third of CrossHair's, mostly input-domain corner cases that inspection overlooks.}
\chgi{Combining agent-written properties~\cite{maaz2025agentic} with the synthesized environment, so that property-based testing can also reach database-backed code, is a natural next step.}

\section*{Acknowledgment}
This work was partially supported by the Wallenberg AI, Autonomous Systems and
Software Program (WASP) funded by the Knut and Alice Wallenberg Foundation.

\bibliographystyle{IEEEtran}
\bibliography{related_works}

\end{document}